\documentclass[aps,twocolumn]{revtex4}

\usepackage{amsfonts}
\usepackage{graphicx}
\usepackage{color}

\begin{document}

\title{Moving boundary approximation for curved streamer ionization fronts: Numerical tests}
\author{Fabian Brau$^1$, Alejandro Luque$^1$, Benny Davidovitch$^{2}$ and Ute Ebert$^{1,3}$ }
\affiliation{$^1$Centrum Wiskunde \& Informatica (CWI), P.O.Box 94079, 1090 GB Amsterdam, The Netherlands}
\affiliation{$^2$Physics Department, University of Massachusetts, Amherst MA 01003} 
\affiliation{$^3$Dept. Physics, Eindhoven Univ. Techn., The Netherlands}

\date{\today}

\begin{abstract}
Recently a moving boundary approximation for the minimal model for negative streamer ionization fronts was extended with effects due to front curvature; this was done through a systematic solvability analysis. A central prediction of this analysis is the existence of a nonvanishing electric field in the streamer interior, whose value is proportional to the front curvature. In this paper we compare this result and other predictions of the solvability analysis with numerical simulations of the minimal model.
\end{abstract}

\maketitle

\section{Introduction}

Streamers characterize the initial stages of electric breakdown in sparks, lightning and sprite discharges; they occur equally in technical and natural processes \cite{raiz91,PSST,ED08}. They are growing plasma channels that appear when strong electric fields are applied to ionizable matter. The essential features of negative (anode-directed) streamers in a non-attaching gas such as argon or nitrogen can be described by the so-called minimal model~\cite{PRLUWC,PREUWC,Kunhardt,dhal85,dhal87,vite94,PRLMan,AndreaRapid,CaroPRE,CaroJCP,li07}.
This model consists of a set of three coupled partial differential equations for the electron density
$\sigma$, the ion density $\rho$ and the electric field ${\bf E}$. In dimensionless units the model reads
\begin{eqnarray}
\label{PDE1}
\partial_t \sigma - \nabla \cdot (\sigma{\bf E})-D\nabla^2 \sigma&=&\sigma|{\bf E}| \;\alpha(|{\bf E}|),
\\
\label{PDE2}
\partial_t \rho &=& \sigma|{\bf E}| \;\alpha(|{\bf E}|),
\\
\label{PDE3}
\nabla \cdot {\bf E}&=&\rho-\sigma, \quad {\bf E} = -\nabla \phi,
\end{eqnarray}
where $D$ is the electron diffusion coefficient and where
\begin{equation}
    \alpha(|{\bf E}|)=e^{-1/|{\bf E}|}.
\end{equation}
A general discussion of the physical dimensions for this model can be found, e.g., in
\cite{PRLUWC,PREUWC,PSST,luqu07}. The model is based on a continuum approximation with local
field-dependent impact ionization reaction. Equations (\ref{PDE1}) and (\ref{PDE2}) are the continuity
equations for the electrons and the ions, taken as immobile due to their much larger mass, while
Eq.~(\ref{PDE3}) is the Coulomb equation for the electric field generated by the space charge $\rho-\sigma$ of electrons and ions. Although discharges in air require extensions of the model, simulation results of negative air streamers frequently resemble the minimal model remarkably well~\cite{luqu07,luqu08}.

\begin{figure}
\centerline{\includegraphics[width=\columnwidth,clip]{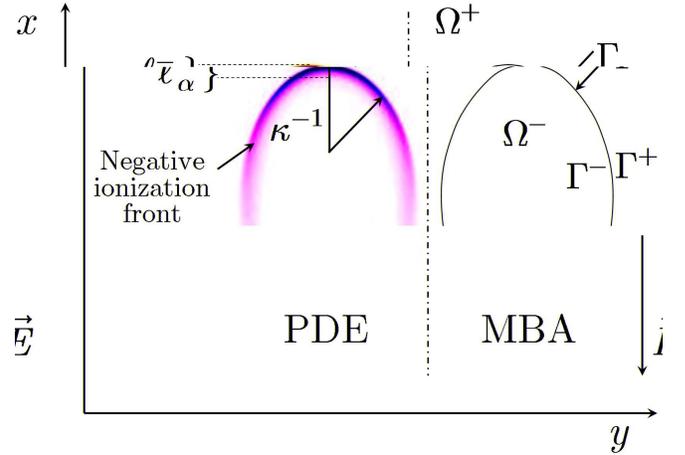}}
\caption{(Color online) On the left: representative solution (net charge density) of the minimal PDE model with curvature $\kappa$ and width $\ell_{\alpha}^-$ (see Eq.~(\ref{ell-al})) of the ionization front. The electric field $\vec{E}$ is pointing downward and the negative front propagates upward. On the right: depiction of the corresponding moving boundary approximation (MBA) with the ionized region, $\Omega^-$, the non-ionized region, $\Omega^+$, and the sharp interface, $\Gamma$.} \label{fig01}
\end{figure}

Many simulations~\cite{Kunhardt,dhal85,dhal87,vite94,PRLMan,AndreaRapid,CaroPRE,CaroJCP} have shown that streamers form a thin curved space charge layer which separates the ionized interior region, $\Omega^-$, from the nonionized exterior region, $\Omega^+$. This narrow charged layer (the ionization front) enhances the electric field in $\Omega^+$ ahead of the front and screens it partially in $\Omega^-$. In strong background fields after some transient evolution, the width of the ionization front can be much smaller than its radius of curvature~\cite{CaroPRE,CaroJCP,luqu07}. This separation of scales enables one to consider the front as an infinitesimally thin, sharp moving interface $\Gamma(t)$. In Fig.~\ref{fig01}, we show a representative snapshot of net charge density of the minimal model (\ref{PDE1})-(\ref{PDE3}) that shows the separation of scales, and depict the corresponding moving boundary approximation. The original nonlinear dynamics is then replaced by a set of linear field equations (frequently of diffusive or Laplace type) on both sides of $\Gamma(t)$; the regions on both sides of $\Gamma(t)$ are denoted as $\Omega^+$ and $\Omega^-$. The linear fields in these regions are determined by boundary conditions on both sides of the interface, $\Gamma(t)^+$, $\Gamma(t)^-$, respectively, and on the outer boundaries (assumed to be located far away from $\Gamma(t)$); the nonlinearity enters through the motion of the boundary. The interface dynamics is typically related to gradients of the Laplacian fields in its vicinity.

In the context of streamer dynamics, the concept of an interfacial approximation was probably first sketched by S\"ammer in 1933 \cite{Saem33}; later it was developed further by Lozansky and Firsov in the Russian literature and in English in~\cite{loza73}. They considered the streamer interior, $\Omega^-$, as ideally conducting, i.e., the electric potential $\phi$ to be constant in the interior. The exterior, $\Omega^+$, is nonionized and therefore does not contain space charges; the electric potential here solves
\begin{equation}
\label{1} \nabla^2\phi=0 \quad \text{in} \quad \Omega^+.
\end{equation}
The interface was assumed to move with the local electron drift velocity
\begin{equation}
\label{2}
{\bf v}= \nabla\phi^+.
\end{equation}
Hence, superscripts $^\pm$ attached to fields, potential and densities indicate their limit values as they approach the interface from $\Omega^+$ and $\Omega^-$, respectively. In particular, we denote $\phi^+\equiv\phi|_{\Gamma^+}$ and $\phi^-\equiv\phi|_{\Gamma^-}$.

However, this simplest moving boundary approximation is mathematically ill posed; in the context of similar models in fluid dynamics, this is explained for example in Ref.~\cite{howi86} and references therein. To resolve this problem, the boundary condition $\phi^+-\phi^-=0$ was replaced by the regularizing boundary condition
\begin{equation}
\label{GLkatze} \phi^+-\phi^-=Q_0(\hat{{\bf n}}\cdot \nabla\phi^+)\stackrel{|\hat{{\bf n}}\cdot
\nabla\phi^+|\gg1}\longrightarrow \hat{{\bf n}}\cdot \nabla\phi^+
\end{equation}
which was proposed in~\cite{PRL05} and derived in planar front approximation in~\cite{brau07}. The boundary condition accounts for the finite width of the charged layer that leads to a finite variation of the electric potential across the front. The boundary condition in the limit of large electric fields actually turns out to be identical to the ``kinetic undercooling'' boundary condition that was applied to crystal growth under certain conditions \cite{kuik85,chap03}. Solutions of the model (\ref{1})--(\ref{GLkatze}) are discussed in~\cite{PRL05,SIAM07}, and the analysis in~\cite{SalehI} shows that the boundary condition (\ref{GLkatze}) indeed regularizes the problem. This moving boundary approximation is compared with solutions of the minimal model (\ref{PDE1})--(\ref{PDE3}) in~\cite{brau07,brau08b}.

In a recent paper~\cite{brau08}, effects associated with curvature of the front were considered. The moving boundary conditions for a slightly curved front dynamics were systematically derived from the original nonlinear field equations (\ref{PDE1})-(\ref{PDE3}), with $D=0$, using the following procedure. A perturbation of a planar front is assumed whose curvature in the direction transverse to the front motion is much smaller than the front width,
\begin{equation}
    \label{eps}
    \epsilon = \ell_{\alpha}^- \kappa \ll 1,
\end{equation}
where $\ell_{\alpha}^-$ and $\kappa$ are respectively the width and the curvature of the front respectively. The width of the front, $\ell_{\alpha}^-$, is taken as the decay length in the ionized region of the net charge density of the planar front with $D=0$ and reads~\cite{brau08}
\begin{equation}
    \label{ell-al}
    \ell_{\alpha}^-(E^+)=\frac{E^+}{\sigma^-(E^+)},
\end{equation}
where $\sigma^-$ is the value of the electron density far behind the planar front whose expression is given by Eq.~(\ref{sigmam}), see for example Refs.~\cite{PREUWC,arra04,brau08}. $\ell_{\alpha}^-(E^+)$ is a monotonically decreasing function of $E^+$ and tends to 1 (in our dimensionless units) for $E^+ \to +\infty$. The computation is carried out in first order in $\epsilon$. Solvability analysis is then used to connect the perturbed values of the fields ahead and behind the curved front to derive the moving boundary approximation. Similarly to other well-studied problems such as solidification dynamics~\cite{karm98,elde01}, this expansion around the planar front solution is asymptotic and does not necessarily converge. However, such solvability analysis provides valuable approximation for the nonlinear dynamics of the propagating front as long as $\epsilon$ remains sufficiently small. Furthermore, notice that such an analysis could not be performed on the streamer model with $D>0$ as the fronts are pulled \cite{eber98, PhysD00, eber00}. However, the leading edge that pulls the front is diffusive, and it is a physically and mathematically meaningful approximation to neglect electron diffusion in strong fields, where electron motion due to drift dominates over the diffusive motion \cite{arra04,derk08}.

The complete model derived in Ref.~\cite{brau08} reads
\begin{eqnarray}
\nabla^2 \phi &=& 0 \quad \text{in} \quad \Omega^+ \label{laplaceahead} \\
\nabla^2 \phi &=& 0 \quad \text{in} \quad \Omega^- \label{laplaceouter}
\end{eqnarray}
with the moving boundary conditions
\begin{eqnarray}
\hat{{\bf n}} \cdot \nabla \phi^- &=& Q_2(\hat{{\bf n}} \cdot \nabla \phi^+)
\kappa \label{BC2} \\
\phi^+-\phi^- &=& Q_0(\hat{{\bf n}} \cdot \nabla \phi^+) + Q_1(\hat{{\bf n}} \cdot \nabla \phi^+) \kappa \label{BC1} \\
v_{n} &=& \hat{{\bf n}} \cdot \nabla\phi^+ \label{velocityfront}.
\end{eqnarray}
The coefficients $Q_i$ depend on the electrostatic field ahead of the front, and are given by analytic
expressions derived from the planar front solution as follows
\begin{eqnarray}
    \label{q0}
    Q_0(y)&=&\int_0^{y} dz\, \frac{y-z}{\rho(z,y)}, \\
    \label{q1}
    Q_1(y)&=&-y\int_0^{y} dx \frac{(y-x)\alpha(x)}{x\rho(x,y)}\int_0^{x}dz
\frac{(y-z)}{\rho(z,y)^2} \nonumber \\ &-& \frac{1}{\sigma^-(y)}\int_0^{y} dx
\frac{(y^2-x^2)}{x\rho(x,y)^2}\left[\rho(x,y)y-\sigma^-(y-x) \right] \nonumber \\
&+&\frac{y}{\sigma^-(y)}\int_0^{y}dx\frac{y-x}{\rho(x,y)}-\frac{y^3}{(\sigma^-(y))^2}, \\
    \label{q2}
    Q_2(y)&=&\frac{y^2}{\sigma^-(y)},
\end{eqnarray}
where
\begin{eqnarray}
    \label{rho}
    \rho(x,y)&=&\int_{|x|}^{y}d\mu \, \alpha(\mu), \\
    \label{sigmam}
    \sigma^-(y)&=&\int_0^{y}d\mu\, \alpha(\mu),
\end{eqnarray}
with $\alpha(x)=e^{-1/x}$. The quantity $\rho$ is related to the ion density profile of the uniformly
translating planar front solution of the minimal model (\ref{PDE1})-(\ref{PDE3}) with $D=0$, see for example Refs.~\cite{PREUWC,arra04,brau08}.

The boundary condition (\ref{BC2}) implies that the electric field just behind the ionization front is not completely screened but that it is proportional to the curvature. This implies that the ideal conductivity approximation in the streamer interior ($\phi=0$ in $\Omega^-$) must be relaxed. In Ref.~\cite{brau08}, the streamer interior was therefore approximated by assuming charge neutrality ($\nabla^2 \phi=0$ in $\Omega^-$). Consequently, the boundary condition (\ref{BC2}) introduces new physics and leads to a new type of moving boundary problem. 

The purpose of this work is to study the validity of this moving boundary model and, for the reason just mentioned, especially the validity of the boundary condition (\ref{BC2}) by comparing it to the results obtained from numerical simulations using the minimal model (\ref{PDE1})-(\ref{PDE3}). Notice that the numerical simulations are performed with a non-vanishing electron diffusion coefficient $D$. There are essentially two reasons for that. First, a vanishing diffusion coefficient leads to a continuum model with an electron discontinuity that cannot be simulated with the numerical methods developed for nonvanishing diffusion. Second, we want to test our boundary conditions on a realistic model and see if our moving boundary model is robust against some changes in the underlying minimal model.

Another relation derived in Ref.~\cite{brau08}, that does not appear explicitly in the model
(\ref{laplaceahead})-(\ref{velocityfront}), can also be tested against numerical simulations. This is the curvature correction to the value of the electron density behind the front
\begin{equation}
    \label{sigma-back}
    \sigma_{\text{back}}=\sigma^-(\hat{{\bf n}} \cdot \nabla \phi^+)+ Q_3(\hat{{\bf n}} \cdot \nabla \phi^+) \kappa,
\end{equation}
where
\begin{equation}
    \label{q3}
    Q_3(y)= y \int_0^y dx \frac{(y-x)\alpha(x)}{x \rho(x,y)}.
\end{equation}

The paper is organized as follows. In Sec.~\ref{sec:comp-bc2-bc1}, we describe the method used for comparing the moving boundary approximation with the simulation data, and in Sec.~\ref{sec:res}, we describe in detail the results of our comparison concerning the boundary conditions (\ref{BC2}) and (\ref{BC1}) and also Eq.~(\ref{sigma-back}). 

\section{Method for comparing the moving boundary approximation with simulations of the minimal model}
\label{sec:comp-bc2-bc1}

In this section, we test the boundary conditions (\ref{BC2}) and (\ref{BC1}) as well as Eq.~(\ref{sigma-back}) against results of simulations of the minimal model (\ref{PDE1})-(\ref{PDE3}) in two dimensions. The electric field and the electron density behind the front are essentially constant over a significant interval, therefore it is relatively easy to extract their values from simulation data without introducing significant errors. The comparison with predictions of Eqs.~(\ref{BC2}) and (\ref{sigma-back}) allow to test the model with confidence. In contrast, as explained below, due to some arbitrariness of the precise location of $\phi^+$ in the simulations and since the potential varies significantly over short distances, the comparison between Eq.~(\ref{BC1}) and the simulations is not quite conclusive.

In order to test our boundary conditions, we need to evaluate the profiles of the net charge density, of electric field and potential and of the electron density along some given axis of the two-dimensional simulated streamer. In this paper we consider a streamer that evolves from initial conditions with a mirror symmetry $y\to -y$. Furthermore, we restrict our analysis to the symmetry axis of the streamer, $y=0$, since this allows easier evaluation of the curvature and the various fields.

\subsection{Numerical simulations}
\label{sec:num-sim}

The minimal PDE model (\ref{PDE1})-(\ref{PDE3}) (with $D=0.1$~\cite{PRLUWC,PREUWC,PSST}) is solved numerically in two dimensions on adaptively refined comoving grids with a second-order explicit Runge-Kutta time integration. The algorithm is
described in detail in Ref.~\cite{CaroJCP} for three-dimensional, cylindrically symmetrical geometries. It is trivially adapted to planar two-dimensional systems, as previously discussed in
Refs.~\cite{brau07,brau08b}. The highest spatial resolution in the area around the streamer head was $\Delta x= \Delta y=1/4$ for all simulations. The simulation domain was $0\le x \le 2048$ and $-1024\le y\le 1024$. The initial conditions were an electrically neutral Gaussian seed of width 16 and height $2.4 \cdot 10^{-5}$. We used four different values for the background electric field applied between the electrodes, namely $E_{\infty}=-0.5, -1, -1.5$ and $-2$. The simulations are the same as in~\cite{brau07}; and for the actual density and field configurations, we refer to the figures in that paper.

\subsection{Extracting relevant quantities from the simulation data} 
\label{procedure}

For each value of $E_{\infty}$, we collected at constant time steps, up to the time of branching, the values of the curvature of the front, of the enhanced electric field (defined as the maximum of the electric field along the symmetry axis) and the profiles of the electric potential and electron density. These are the ingredients of the boundary conditions (\ref{BC2}) and (\ref{BC1}) and of Eq.~(\ref{sigma-back}) that we test in this paper.

\begin{figure}
\centerline{\includegraphics[width=\columnwidth,clip]{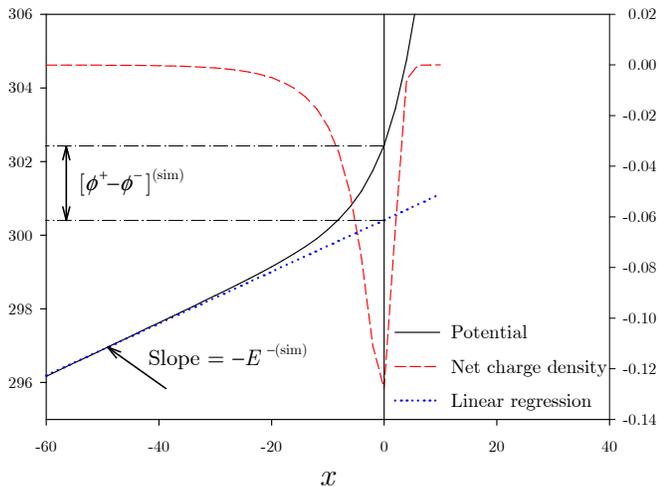}}
\caption{(Color online) Profiles of the electric potential and of the negative net charge density along the symmetry axis, $y=0$, for $E_{\infty}=-0.5$ and $t=490$. The linear regression for the linear part of the potential is also plotted. The slope of this linear part corresponds to the electric behind the front, $E^{-(\text{sim})}$. The difference between the simulated value of the potential at $x=0$ and the value of the extrapolation of the linear part at the same location gives the jump in the electric potential.}
\label{figure02}
\end{figure}

To test both relations (\ref{BC2}) and (\ref{BC1}) using a unique procedure, we consider Eq.~(128) of
Ref.~\cite{brau08} (where the leading contribution of the planar front is added here),
\begin{equation}
    \label{eqtotest}
    \phi(0)-\phi(x) = Q_0(E^+) + Q_1(E^+)\kappa - Q_2(E^+) \kappa\, x,
\end{equation}
where $|x| \gg \ell^-_{\alpha}$, and where $x=0$ corresponds to the position of the tip of the front, i.e. the position of the discontinuity line $\Gamma(t)$. This equation predicts that the correction to the potential profile due to curvature is a linear function of the variable $x$ behind the front in an intermediate region between the inner region and the outer region~\cite{brau08}. The position $x=0$ where the potential $\phi(0)$ is evaluated, is taken as the location of the maximum of the negative net charge density, while $E^+$ is identified with the maximum of the electric field along the symmetry axis (since we restrict our analysis here to this axis). The slope of the linear curve (\ref{eqtotest}) yields the boundary condition (\ref{BC2}) and expresses the electric field behind the front. The comparison between the slope in the simulations and the corresponding expression in Eq.~(\ref{eqtotest}) is then a direct test of Eq.~(\ref{BC2}). Equation (\ref{eqtotest}) can also be used to test the relation (\ref{BC1}). For this purpose, we extrapolate the linear part of the potential, $\phi_{\text{lin}}$, up to the tip of the front ($x=0$) and compare the difference between this value obtained for the potential, $\phi_{\text{lin}}(0)$, and the value of the potential obtained from the simulation at $x=0$, $\phi(0)$. Indeed we have
\begin{equation}
    \label{eqtotest2}
    \phi(0)-\phi_{\text{lin}}(0) = Q_0(E^+) + Q_1(E^+)\kappa,
\end{equation}
where $\phi(0)$ can be measured from the simulated potential and $\phi_{\text{lin}}(0)$ is obtained from the
linear regression; the procedure is illustrated for $E_{\infty}=-0.5$ and $t=490$ in Fig.~\ref{figure02}. We then can compare the simulated potential jump with the theoretical value on the right-hand side of Eq.~(\ref{eqtotest2}) which corresponds to the boundary condition (\ref{BC1}).

In order to compute the curvature of the ionization front, we need to define a one-dimensional curve from the diffuse two-dimensional front. For this purpose, we use the following procedure. Let the streamer propagate along the $x$-axis, $y$ being the transverse axis. For a given value of $x$, we locate the position of the maximum of the net charge density along the $y$-axis to get two points (due to mirror symmetry) of the one-dimensional curve. We repeat the procedure for each value $x$ along the streamer length to get the complete one-dimensional curve: $y(x)$ indicates the position of the maximal charge density for every $x$. The same procedure was used previously in Ref.~\cite{brau08b}. We estimate the curvature $\kappa$ of the front by fitting the section of the curve $y(x)$ with a polynomial $x = - \kappa y^2 / 2+ O(y^4)$.

The enhanced field $E^{+(\text{sim})}$ is identified with the maximum of the absolute value of the electric field in
the simulations (along the symmetry axis $y=0$). 

The extraction of $E^-$ from the simulations, $E^{-(\text{sim})}$, is obtained from the profile of the
electric potential along the symmetry axis as already explained above (see also Fig.~\ref{figure02}) while its value obtained within the moving boundary approximation, $E^{-(\text{MBA})}$, is computed using Eq.~(\ref{BC2}).

The potential behind the front, $\phi^-$, is obtained together with $E^{-(\text{sim})}$ since the latter is given by the slope of the linear part of the simulated potential behind the front while the former is given by the intersection of the linear regression with the position of the tip of the front (the position of the discontinuity line $\Gamma(t)$) that here was chosen to be the maximum of the net charge density.

The potential ahead of the front, $\phi^+$, is identified with the value of the potential at the location of the maximum of the net charge density. We also report later in Fig.~\ref{figure08} the values of the potential at two grid points on our finest grid, adjacent to the location chosen to be the discontinuity line, $\Gamma(t)$, of the front. 

\begin{figure}
\centerline{\includegraphics[width=\columnwidth,clip]{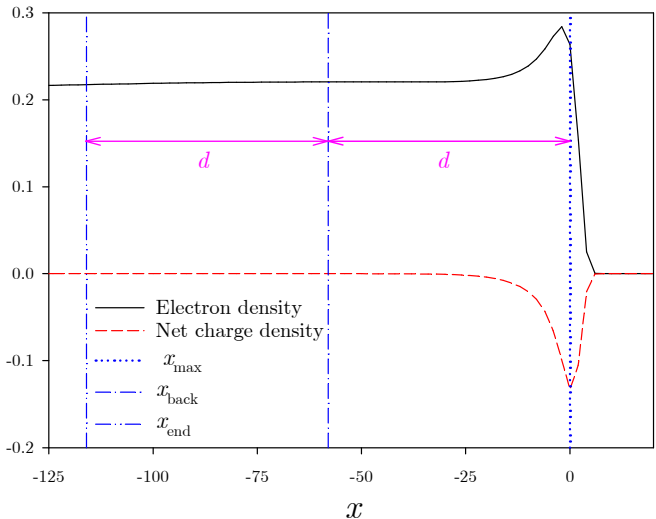}}
\caption{(Color online) Electron and net charge density profiles for $E_{\infty}=-0.5$ and $t=490$. The position of the maximum of the negative net charge density, $x_{\text{max}}$, the position where the charge density vanishes, $x_{\text{back}}$, and $x_{\text{end}}$ (see text) are also indicated. The distance $d$ is equal to $x_{\text{max}} - x_{\text{back}}$.} \label{figure03}
\end{figure}

The electron density behind the negative front, $\sigma_{\text{back}}^{(\text{sim})}$, is obtained from the simulations as $\sigma_{\text{back}} = \frac{1}{2}(\sigma(x_{\text{back}}) +
\sigma(x_{\text{end}}))$, where $x_{\text{back}}$ is the position where the net charge density vanishes. Such point must exist since we start with a neutral seed between the electrodes and thus when the streamer forms, positive and negative net charge densities form at the streamer edges and, consequently, the charge density vanishes somewhere in between. The abscissa $x_{\text{end}}$ is defined as
$x_{\text{end}}=x_{\text{max}}-2(x_{\text{max}}-x_{\text{back}})$, where $x_{\text{max}}$ is the position of the maximum of the net charge density, see Fig.~\ref{figure03}. The quantity $\sigma(x_{\text{end}})$ ($\sigma(x_{\text{back}})$) is then the lower (upper) end of the error bars on the value of the electron density behind the negative front. This procedure gives an estimation of the interval of variation of $\sigma$ behind the front. In Fig.~\ref{figure03}, we illustrate the procedure for $E_{\infty} = -0.5$ at time $t = 490$. 

The main source of errors in extracting the relevant quantities from the simulation data is the diffusive nature of the simulated front and hence the non-uniqueness in identifying the interface $\Gamma(t)$. This uncertainty has no influence on the extraction of the quantities $E^{-(\text{sim})}$ and $\sigma_{\text{back}}^{(\text{sim})}$ from the simulation data since those quantities are evaluated far enough behind the front where they are essentially constant. The error on the slope of the linear part of the potential behind the front, which gives $E^{-(\text{sim})}$, is negligible for our purpose. The interval $[x_{\text{end}},x_{\text{back}}]$ where we chose to measure $\sigma_{\text{back}}^{(\text{sim})}$ is, of course, somewhat arbitrary but since the electron density is essentially constant behind the front, another procedure would give equivalent results; only the size of the error bars could be slightly different. Consequently the errors on these two quantities are well controlled. The errors for the extracted value $E^{+(\text{sim})}$ are also negligible for our purpose since this quantity is evaluated on the finest grid ($\Delta x=\Delta y = 1/4$) used in the simulations. However, the uncertainty about the exact position of $\Gamma(t)$ affects the extraction of $\phi^+$ and $\phi^-$. Indeed, the location where we chose to evaluate $\phi^+$ and $\phi^-$ on the potential profile is rather arbitrary. Moreover, the potential and the linear regression vary significantly over short distances as shown in Fig.~\ref{figure02}. Consequently, the uncertainty of the location of $\phi^+$ (and thus of $\phi^-$) directly influences the results of the comparison between the moving boundary approximation and simulations. However, as explained in Sec.~\ref{sec:test-phi}, the value $[\phi^+ - \phi^-]^{(\text{sim})}$ extracted from the simulation data using the procedure described above is an upper bound on the actual value of the potential jump.

\subsection{Influence of the background electric field}

\begin{figure}[!hbtp]
\centerline{\includegraphics[width=\columnwidth,clip]{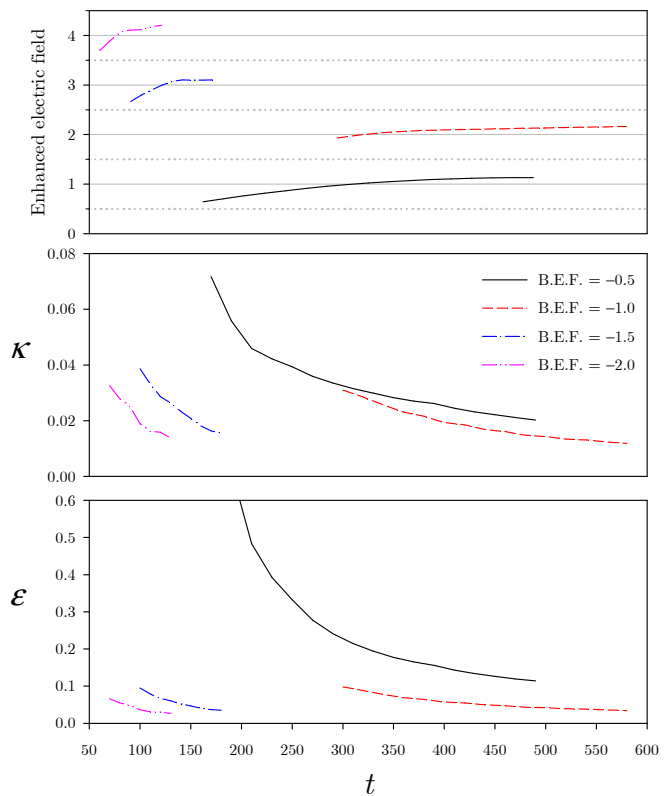}}
\caption{(Color online) Top: Evolution of the absolute value of the maximum of the electric field along the symmetry axis, $y=0$, as a function of time for four values of the background electric field $E_{\infty}$ (B.E.F. on the figure). Middle: Evolution of the curvature of the tip of the front as a function of time for the same values of $E_{\infty}$. Bottom: $\epsilon$ as a function of time for the same values of $E_{\infty}$.} \label{figure04}
\end{figure}

We expect that the simulation results are better approximated by the moving boundary approximation
(\ref{BC2}), (\ref{BC1}) and (\ref{sigma-back}) when the background electric field, $E_{\infty}$, is large enough. This is so, since as mentioned above, our boundary conditions are derived in the regime $\ell_{\alpha}^- \kappa \ll 1$. The formula (\ref{ell-al}) and the simulations indicate that the width of the front is controlled by the value of the enhanced field at the tip of the streamer. The formula (\ref{ell-al}) derived for planar fronts catches qualitatively the evolution of
the front width for a planar interface: The width decreases when the enhanced field increases. Moreover, we notice that in the present simulations (in 2D and in a homogeneous electric field) after some initial transients, the value of the enhanced electric field, up to the time of branching, in good approximation is given by (see Fig.~\ref{figure04})
\begin{equation}
    \label{scalinge+}
    |E^+|=2 |E_{\infty}| + \text{small corrections}.
\end{equation}
Consequently, the width of the front is controlled essentially by the background electric field (plus some corrections) and thus for low $E_{\infty}$, where the front width $\ell_{\alpha}^-$ diverges, one can expect that the boundary conditions (\ref{BC2}) and (\ref{BC1}) and Eq.~(\ref{sigma-back}) will not approximate the simulations very well. We show below, however, that for $E_{\infty}\ge 1$, our analytical approximation for the value of the electric field behind the front fits the simulations very well.

For sufficiently large $E_{\infty}$, we expect that the relations (\ref{BC2}), (\ref{BC1}) and
(\ref{sigma-back}) approximate the simulations well, and we also expect that the approximation improves with time. Indeed, starting from the initial neutral seed, first an interfacial layer forms, and then the value of the enhanced field grows and approaches a plateau value given by (\ref{scalinge+}). From Eq.~(\ref{ell-al}), this also means that the width of the front decreases during this time. On the other hand, during this process, the curvature of the front decreases also (see Fig.~\ref{figure04}). Consequently, for a given $E_{\infty}$, the product $\ell_{\alpha}^- \kappa$ decreases during the evolution of the streamer; see Fig.~\ref{figure04}. Consequently, since our boundary conditions are derived for $\ell_{\alpha}^- \kappa \ll 1$, we expect better agreement between the moving boundary approximation and simulations for time and background electric fields large enough.

This discussion is summarized in the lower panel of Fig.~\ref{figure04}, where we show that
$\epsilon=\ell_{\alpha}^- \kappa$ is a decreasing function of time and of $E_{\infty}$.

\section{Results of the comparison}
\label{sec:res}

\subsection{Testing the boundary condition for $E^-$}
\label{sec:test-e}

\begin{figure}
\centerline{\includegraphics[width=\columnwidth,clip]{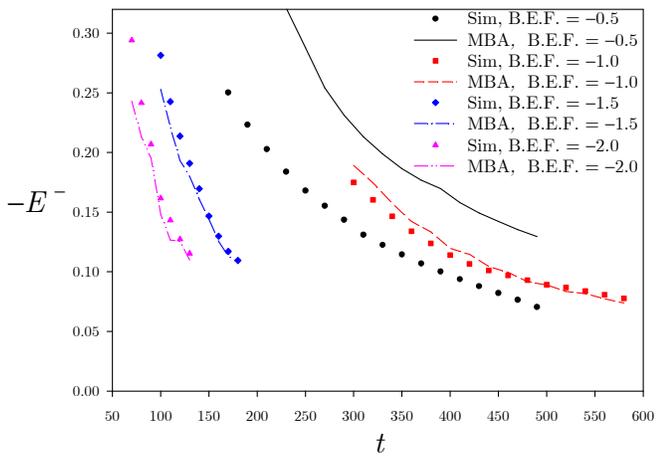}}
\caption{(Color online) Comparison between
the simulated electric field behind negative ionization fronts, $E^{-(\text{sim})}$, and the
values, $E^{-(\text{MBA})}$, computed with the boundary condition (\ref{BC2}) using curvature and
enhanced field, $E^+$, from the simulations. The comparison is performed for four values of
the background electric field (B.E.F. in the figure).} \label{figure05}
\end{figure}

Following the procedures described in Sec.~\ref{procedure}, we extracted the values of $E^{-(\text{sim})}$ from the simulations for four background electric fields: $E_{\infty}=-0.5, -1.0, -1.5$ and $-2.0$. These values are then compared with the values, $E^{-(\text{MBA})}$, predicted by Eq.~(\ref{BC2}) where the curvature, $\kappa$, and the enhanced field , $E^+$, are also obtained from the simulations. The results are compared in Fig.~\ref{figure05}. The error bars of $E^{-(\text{sim})}$ are too small to be visible in the figure.

The agreement between simulations and the moving boundary approximation is rather remarkable except for $E_{\infty}=-0.5$. For this case, the relative errors are always larger than $65 \%$, while for larger background field the errors stay always below $10-12 \%$. In order to understand why the agreement is less good for $E_{\infty}=-0.5$, we compute $\epsilon$ from Eq.~(\ref{eps}). Indeed, we recall that the moving boundary approximation was derived through first order perturbation theory in $\epsilon$. However, the theory, being linear in $\epsilon$, does not provide an estimation for how small $\epsilon$ should be. Fig.~\ref{figure04} shows that for $E_{\infty}=-0.5$, the value of $\epsilon$ stays always above 0.1. Actually from that figure, we can infer that for $\epsilon \lesssim 0.05$, the boundary condition (\ref{BC2}) is accurate within $5 \%$ or less. 

However, at first sight, $\epsilon$ seems not to be the only control parameter. Indeed, for $E_{\infty}=-1.5$ and $t=100$, we read from Fig.~\ref{figure04} that $\epsilon \simeq 0.10$ and we find that the relative error for $E^-$ is about $11 \%$ (see Fig.~\ref{figure05}) while for $E_{\infty}=-0.5$ and $t=490$, we find that $\epsilon \simeq 0.11$ and that the relative error is about $84 \%$. This means that for the same value of $\epsilon$ we get quite different relative errors for the values of the electric field behind the negative front. However for such a value of $\epsilon$, second order terms, neglected in the derivation of the moving boundary approximation (\ref{laplaceahead})-(\ref{velocityfront}), see Ref.~\cite{brau08}, could still play a role. For example, a coefficient associated with $\epsilon^2$ which would decrease fast enough with an increase of the enhanced
field may explain why second order terms are, in this situation, negligible for larger fields while they still play some role for weaker ones. Second order terms could also depend more significantly on the geometry of the streamer by involving a tangential derivative of the curvature. However, without deriving the second order theory, we cannot draw definitive conclusions on this particular issue.

\subsection{Testing the relation for $\sigma_{\text{back}}$}
\label{sec:test-sigma}

\begin{figure}
\centerline{\includegraphics[width=\columnwidth,clip]{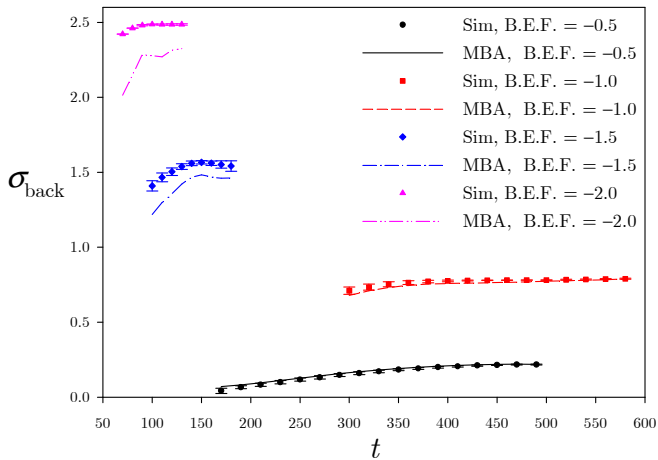}}
\caption{(Color online) Comparison between the value of the simulated electron densities behind the front, $\sigma_{\text{back}}^{(\text{sim})}$, and the values, $\sigma_{\text{back}}^{(\text{MBA})}$,
computed with Eq.~(\ref{sigma-back}) using curvature and enhanced field, $E^+$, of the simulations.
This comparison is performed for four values of the background electric field (B.E.F. in
the figure).} \label{figure06}
\end{figure}

Using the procedure described in Sec.~\ref{procedure}, we estimated the values of
$\sigma_{\text{back}}^{(\text{sim})}$ from the simulations for the same four background electric fields. These values are then compared with the values, $\sigma_{\text{back}}^{(\text{MBA})}$, predicted by Eq.~(\ref{sigma-back}). The results are compared in Fig.~\ref{figure06}.

\begin{figure}
\centerline{\includegraphics[width=\columnwidth,clip]{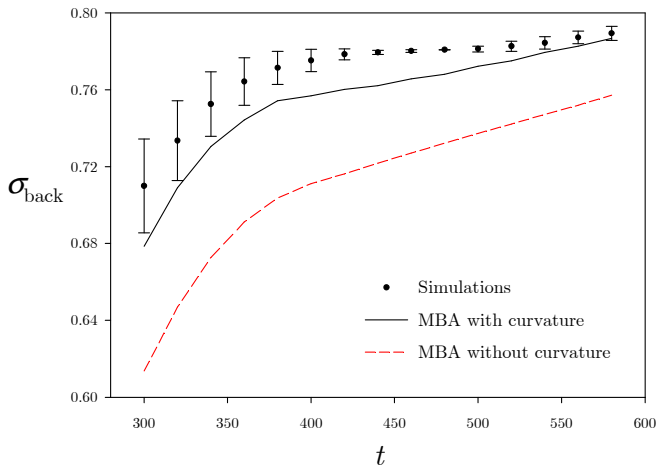}}
\caption{(Color online) Comparison between the values of the simulated electron density behind the front, $\sigma_{\text{back}}^{(\text{sim})}$, the values, $\sigma_{\text{back}}^{(\text{MBA})}$,
computed with Eq.~(\ref{sigma-back}) and the values, $\sigma_{\text{back}}^{(\text{MBA})}$, computed with $\kappa=0$. This comparison is performed for $E_{\infty}=-1.0$.} \label{figure07}
\end{figure}

The simulation values and those of the moving boundary approximation agree rather well.
However, the value of $\sigma_{\text{back}}^{(\text{sim})}$ is slightly underestimated in
larger fields. Nevertheless, for $\epsilon \lesssim 0.05$, the relative errors are about 10$\%$ or less. Moreover, the curvature correction improves the approximation of the electron density behind the front since the additional term is positive (see Eq.~(\ref{sigma-back})). In Fig.~\ref{figure07}, we compare the effects of the curvature correction for $E_{\infty}=-1.0$.

\subsection{Testing the boundary condition for $\phi^+ - \phi^-$}
\label{sec:test-phi}

\begin{figure}
\centerline{\includegraphics[width=\columnwidth,clip]{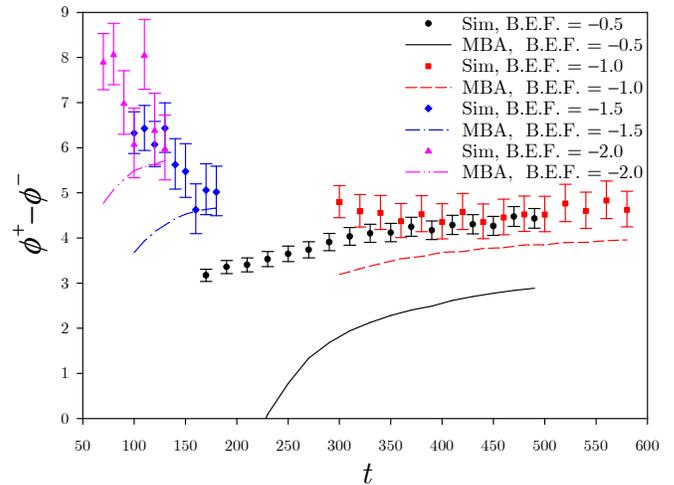}}
\caption{(Color online) Comparison between the jump of the electric potential across the interface
from simulations, $[\phi^+ - \phi^-]^{(\text{sim})}$, and the values, $[\phi^+ - \phi^-]^{(\text{MBA})}$, computed with the boundary condition (\ref{BC1}) where the curvature and the enhanced field, $E^+$, are also obtained from the simulations. This comparison is performed for four values of the background electric field
(B.E.F. in the figure).} \label{figure08}
\end{figure}

Following the procedures described in Sec.~\ref{procedure}, we estimated the values of $[\phi^+ -
\phi^-]^{(\text{sim})}$ from the simulations for the same four background electric fields. These values are then compared with the values, $[\phi^+ - \phi^-]^{(\text{MBA})}$, predicted by Eq.~(\ref{BC1}) (or equivalently Eq.~(\ref{eqtotest2})). The results are compared in Fig.~\ref{figure08}. The lower (upper) end of the error bars for the simulation results correspond to the value of the potential at the grid point just before (after) the position of the maximum of the net charge density on our finest grid. The size of the error bars indicates clearly that indeed the potential varies significantly over quite short distances.

The agreement between the simulation results and the moving boundary approximation for the potential gap, Eq.~(\ref{BC1}), is less satisfactory than the excellent agreement demonstrated above for the electric field and charge density, Eqs.~(\ref{BC2}) and (\ref{sigma-back}). Indeed, for $\epsilon \lesssim 0.05$, the relative error
is about 20$\%$ or less while for $E^-$ and the same values of $\epsilon$, the relative error was about 5$\%$ or less. One reason could simply be that the moving boundary approximation works less well for the jump of the electric potential than for $E^-$, perhaps due to corrections associated with higher order terms in $\epsilon$. However, another reason is certainly that in this analysis there is one arbitrariness: The precise location for evaluation of $\phi^+$. Indeed, as already mentioned above, we choose the location of $\phi^+$ as the location of the maximum of the negative net charge density. Even if this is a rather natural choice, the actual position of $\phi^+$, assumed by the moving boundary approach, could be different. However, because the potential varies significantly over short distances, see Fig.~\ref{figure02} and the size of the error bars on Fig.~\ref{figure08}, the arbitrariness of the location of $\phi^+$ has certainly a direct influence on the comparison between the moving boundary approximation and simulations. For example, another possible location for $\phi^+$ could be the place, $\bar{x}$, such that the amount of negative charge on $x<\bar{x}$ equals the amount of negative charge on $x>\bar{x}$. Since the profile of the net charge density is asymmetric with respect to the position of the maximum (see Fig.~\ref{figure02}), $\bar{x}$ would be located before the position of the maximum ($\bar{x}< x_{\text{max}}$) and the jump of the electric potential extracted from the data would be smaller since the potential and its linear regression are increasing functions of $x$. Consequently, the  
quantity $[\phi^+ - \phi^-]^{(\text{sim})}$ extracted from the data using our procedure is actually an upper bound on the potential jump assumed in the moving boundary approach.

\section{Conclusions}

In this paper, we have tested the recently derived moving boundary approximation~\cite{brau08} for negative ionization fronts on simulations of the minimal model (\ref{PDE1})-(\ref{PDE3}). For this purpose, we employed a simple geometry, characterized by mirror symmetry, and performed our analysis only along the symmetry axis. Future work may seek to perform similar analysis away from the symmetry axis. Our analysis confirmed the validity of two out of the three moving boundary conditions derived in~\cite{brau08}, pertaining to the curvature dependence of the electrostatic field and the charge density in the ionized region behind the propagating front. We showed that these boundary conditions are satisfied for slightly curved fronts, characterized by a small ratio between the front width $\ell^-_{\alpha}$ and the radius of curvature $\kappa^{-1}$ of the front. A third boundary condition, concerning the potential jump across the curved front has not been fully confirmed -– a problem that we attribute to the inherent arbitrariness in extracting from simulations the appropriate potential values (corresponding to their value at the discontinuity line assumed by the moving boundary approach). Further study of the range of validity of this condition will require the development of quantitative tools for such analysis.

Finally, the usefulness of the moving boundary approach for analytic and numerical studies of streamer dynamics depends crucially on its capability to describe front dynamics when the ratio $\epsilon$ is not small, as could happen, at least in principle, along some regimes of the propagating front. A progress in studying this important question will require extension of the MBA derived in~\cite{brau08} to such regime, and comparison with numerical simulations along the approach developed in this paper.

\begin{acknowledgments}
The work of F.B. was supported by The Netherlands Organization for Scientific Research (NWO) through
Contract No. 633.000.401 within the program ``Dynamics of Patterns''. The work of A.L. was supported by the Dutch STW project 06501. The work of B.D. in The Netherlands was supported by a visitor's grant of NWO.
\end{acknowledgments}

\end{document}